\documentclass{svjour3}
\usepackage{epsfig}

\begin{document}

\title{Analytical study of anisotropic compact star models}
\author{B. V. Ivanov}
\institute{B.V.Ivanov\at
Institute for Nuclear Research and Nuclear Energy, \\
Bulgarian Academy of Science, \\
Tzarigradsko Shausse 72, Sofia 1784, Bulgaria\\
\email{boykovi@gmail.com}}
\maketitle

\begin{abstract}
A simple classification is given of the anisotropic relativistic star
models, resembling the one of charged isotropic solutions. On the ground of
this database, and taking into account the conditions for physically
realistic star models, a method is proposed for generating all such
solutions. It is based on the energy density and the radial pressure as
seeding functions. Numerous relations between the realistic conditions are
found and the need for a graphic proof is reduced just to one pair of
inequalities. This general formalism is illustrated with an example of a
class of solutions with linear equation of state and simple energy density.
It is found that the solutions depend on three free constants and concrete
examples are given. Some other popular models are studied with the same
method.
\end{abstract}

\section{Introduction}

The study of relativistic stellar structure is now more than 100 years old.
It began with the discovery in 1916 by Karl Schwarzschild of a universal
vacuum exterior solution \cite{one}. He also gave in the same year the first
interior stellar solution \cite{two}, which should be matched to the
exterior one. It has a constant energy density $\rho $. For a long time the
star interior was considered to be made of perfect fluid, which has equal
radial ($p_r$) and tangential ($p_t$) pressures. This leads to the isotropic
condition $p_r=p_t$, imposed on the Einstein equations. However, spherical
symmetry demands only the equality of the two tangential pressures. This
fact was noticed by Einstein and developed first by Lemaitre in 1933 \cite
{three}. He discussed a model sustained solely by $p_t$ and with constant $%
\rho $. His work remained unnoticed for a long time.

In 1972 Ruderman \cite{four} argued for the first time that nuclear matter
at very high densities of the order of $10^{15}$ $g/cm^3$ may have
anisotropic features and its interactions are relativistic. The pioneering
work of Bowers and Liang \cite{five} on building anisotropic models in 1974
gave start to a number of such solutions. Anisotropy may have a lot of
sources \cite{six}: a mixture of fluids of different types, presence of a
superfluid, existence of a solid core, phase transitions, presence of
magnetic field, viscosity, etc. Such models describe compact stellar objects
like neutron stars, strange stars, quark stars, boson stars, gravastars,
dark stars and others.

The Einstein equations describe the effect of matter upon the metric of
spacetime. For static, spherically symmetric fluid solutions the metric may
be written in canonical or isotropic coordinates and has two components $\nu 
$ and $\lambda $ or $\nu $ and $\mu $. The energy-momentum tensor is
represented by its diagonal components, mentioned above: $\rho $, $p_r$ and $%
p_t$. There are only three equations for these five characteristics, so that
two of them may be chosen freely. They should satisfy, however, a lot of
regularity, stability and energy conditions for a realistic model. The
situation is analogous to the search of charged isotropic star models \cite
{seven}. This is not surprising since charge can be looked upon as an
effective anisotropy of the model \cite{eight}. Different choices of the two
given functions have been made.

The simplest one is to propose ansatze for the two metric functions. Then
the Einstein equations become expressions for the matter components. One of
the first solutions was given in \cite{nine}, where some of the Tolman
isotropic solutions \cite{ten} were modified to become anisotropic. Another 
\cite{eleven} relies on the metric of the well known charged Krori and Barua
solution \cite{twelve}. Others were based on different isotropic solutions,
found in the past \cite{thirteen}, \cite{fourteen}. A solution in isotropic
coordinates also exists \cite{fifteen}.

String theory has inspired embedding of branes like in the Randall - Sundrum
model \cite{sixteen}. This rekindled the interest in stellar models embedded
in five-dimensional flat spacetime (embedding class one). They must satisfy
the Karmarkar condition \cite{seventeen}. It can be written as a relation
between $\lambda $ and $\nu ^{\prime }$ where $^{\prime }$ means radial
derivative. Consequently, one can choose one of these functions as
generating the whole solution. It is interesting that the isotropic
condition, which is an equality between the pressures, can be translated
into a similar relation, giving different generating functions \cite
{eighteen}, \cite{nineteen}, \cite{twenty}, \cite{twone}, \cite{twtwo}. This
is easier done in isotropic coordinates, but canonical coordinates can be
used too \cite{twenty}.

There are just two perfect fluid solutions of the Karmarkar condition - the
interior Schwarzschild one \cite{two}, which has infinite speed of sound and
a cosmological one. When the fluid is anisotropic, a plethora of realistic
solutions have been found in the last two years. In some of them the
generating metric component is a polynomial \cite{twthree}, \cite{twfour}, 
\cite{twfive}, \cite{twsix}, \cite{twseven}, \cite{tweight}, in others it is
a rational function \cite{twnine}, \cite{thirty}, \cite{thone}. There are
also trigonometric \cite{thtwo}, \cite{ththree}, hyperbolic \cite{thfour}
and exponential generating metric components \cite{thfive}, \cite{thsix}.

There is a group of hybrid solutions with given one metric and one matter
function. Such is the model in higher dimensional spacetime with prescribed $%
\nu $ and the anisotropy factor $\Delta =p_t-p_r$, which measures the
diversion from isotropy \cite{thseven}. An algorithm was given how to find
any anisotropic solution provided these potentials are known \cite{theight}.
It is based on linear differential equations, which are integrable.
Different examples were given, but the regularity conditions were not
studied. For isotropic solutions it becomes the algorithm of Lake \cite
{twenty}. It was used in \cite{thnine}, where a special function appears.
Solutions were also found in isotropic coordinates \cite{forty}, \cite{foone}

Conformally flat anisotropic spheres have vanishing Weyl tensor. This leads
to a linear differential equation for $\nu ^{\prime }$, which can be
integrated and even a relation between $\nu $ and $\lambda $ is the outcome 
\cite{fotwo}. Several models were given with $p_r=0$ or prescribed $\lambda $%
. Recently, a conformally flat model with polytropic equation of state was
discussed \cite{fothree}. More solutions have been given by other authors 
\cite{fofour}, \cite{fofive}.

Closely related are solutions which admit conformal motion. They depend on
the conformal factor and a matter component, which can be $\rho $ \cite
{fosix}, \cite{foseven}, or the mass $m$ \cite{foeight}. One can add here a
model with given $\lambda $ \cite{fonine}, since the expressions for $\rho $%
, $m$ and $\lambda $ are simply related. There is a model with a linear
equation of state (LEOS) between $p_r$ and $\rho $ \cite{fifty}, and another
one with LEOS between the pressures \cite{fione}. The work \cite{fotwo} has
been generalized to non-static and electrically charged solutions \cite
{fitwo}.

Algorithms have been proposed for obtaining anisotropic solutions from
isotropic relativistic \cite{fithree}, \cite{fifour}, \cite{fifive} or
Newtonian ones \cite{fisix}.

The last group of known solutions are the models with two freely prescribed
matter components. The interior Schwarzschild solution \cite{two} is one of
them, but it is isotropic with $p_r=p_t$ and $\rho =const$. The first
anisotropic solution was proposed by Lemaitre \cite{three}. It has vanishing 
$p_r$ and constant $\rho $ and was found independently by Florides \cite
{fiseven}. It was studied further in \cite{fieight}. A very important
equation of hydrostatic equilibrium exists, involving only $\rho $, $p_r$
and $p_t$ in canonical coordinates. This is the TOV (Tolman, Oppenheimer and
Volkoff) equation \cite{ten}, \cite{finine} found initially for isotropic
solutions. Its anisotropic version was used by Bowers and Liang \cite{five}
to find the first well-known star model, which has constant $\rho $, while $%
\Delta $ is given in a form suitable to solve the TOV equation. They
asserted that anisotropic models may have arbitrarily large surface
redshift, which can explain the big redshifts of quasars. However, when the
energy conditions are taken into account, realistic models have bounded
redshift \cite{sixty}. The surface redshift, the mass and the radius of the
star are characteristics that can be measured by astronomers. A model with
constant $\rho $ and non-vanishing $p_r$ was given \cite{sione}. In \cite
{sitwo} the method of \cite{five} was further developed and a model with
singular $\rho $ was studied too. Bondi \cite{sithree} searched for models
with large redshift, including a solution with constant $\rho $ and another,
with constant $Q=p_r+2p_t$, and found that the latter is more perspective.

Recently, it was shown that in isotropic coordinates the existence of an EOS
gives an expression of $\lambda $ in terms of $\mu $, which may serve as a
generating function. The case with $p_r=0$ was solved completely \cite
{sifour}, giving as example a new solution. The same was done for solutions
with LEOS $p_r=\alpha \rho -\beta $ \cite{sifive} and for the Chaplygin EOS 
\cite{sisix}.

A number of solutions with given $\rho $ and $p_r$, not linked by an EOS, is
known \cite{siseven}, \cite{sieight}, \cite{sinine}, \cite{seventy}.
Recently, solutions with prescribed $\lambda $ and $p_r$ have been given
full physical analysis \cite{seone}, \cite{setwo}, \cite{sethree}, \cite
{sefour}. At first sight, these are hybrid solutions, but since $\lambda $
is closely related to $\rho $, we mention them here. Solutions, containing
free $\Delta $, exist in several combinations. Thus $\Delta $, $\rho $
solutions either simplify the TOV equation \cite{sefive}, \cite{sesix} or
the Einstein equations, which acquire simple solutions \cite{seseven}, \cite
{seeight}, or even such in hypergeometric functions \cite{senine}, \cite
{eighty}.

A subgroup of this last group are models, where $p_r$ satisfies some EOS.
The simplest one is the so-called $\gamma $-law, $p_r=\gamma \rho $, linear
and without a free term. Anisotropy allows much more solutions. This
resembles the addition of charge to isotropic solutions \cite{seven}, \cite
{eione}. There are models with prescribed $\rho $ \cite{eitwo}, \cite
{eithree} or $m$ \cite{eifour}. For usual compact stars $\gamma \in [0,1]$.
For dark energy stars it may be negative \cite{eifive}, \cite{eisix}.
Quintessence stars have a second dark energy-density $\rho _q$ imposed on
normal matter \cite{eiseven}, \cite{eieight}, \cite{einine}, \cite{ninety}.
Some earlier models of this type with $p_r=-\rho $ are \cite{nione}, \cite
{nitwo}.

Another class of models have a linear EOS with a free term, the so-called
MIT bag constant, which is suitable for more compact stars, $p_r=\alpha \rho
-\beta $. There are models with given in addition $m$, \cite{nithree}, \cite
{nifour} or $\lambda $ \cite{nifive}, \cite{nisix}, \cite{niseven}, \cite
{nieight}, \cite{ninine}. There are also models with quadratic EOS and $%
\lambda $ \cite{hundred}, \cite{huone}, \cite{hutwo}. The popular from
Newtonian gravity polytropic EOS was shown to lead in the anisotropic
Newtonian case to the well known Lane - Emden equations \cite{huthree} and
to their relativistic generalization in Einstein's gravity \cite{hufour}. A
simple solution for a simple ansatz for $\lambda $ was found \cite{hufive}.
Finally a solution with $\lambda $ and the modified Van der Vaals EOS was
presented recently \cite{husix}.

The above classification of anisotropic star models does not pretend to be
exhaustive, especially for the more exotic cases. There are additional
references, cited in the recent papers on this topic. We wanted to draw a
global picture, showing where does the general method for finding physically
realistic solutions, proposed in the present paper, stand. In Sect. 2 the
Einstein field equations are given, as well as the definitions of the main
characteristics of a static anisotropic star. In Sect. 3 we summarize the
conditions for a physically realistic model, amassed during the past
decades. In Sect. 4 we argue that the model of type $\rho $, $p_r$ is the
easiest one to implement these conditions. In Sect. 5 we derive different
relations between the conditions, which reduce their number. In Sect. 6 an
example is given - solutions with linear EOS and simple energy density. In
this method the main object of study is the tangential pressure, which is
done in Sect. 7. In Sect. 8 some other EOS are studied. Sect. 9 contains
discussion.

\section{Field equations and definitions}

The interior of static spherically symmetric stars is described by the
canonical line element 
\begin{equation}
ds^2=e^\nu c^2dt^2-e^\lambda dr^2-r^2\left( d\theta ^2+\sin ^2\theta
d\varphi ^2\right) ,  \label{one}
\end{equation}
where $\lambda $ and $\nu $ are dimensionless and depend only on the radial
coordinate $r$. The Einstein equations read 
\begin{equation}
k\rho =\frac 1{r^2}\left[ r\left( 1-e^{-\lambda }\right) \right] ^{\prime },
\label{two}
\end{equation}

\begin{equation}
kp_r=-\frac 1{r^2}\left( 1-e^{-\lambda }\right) +\frac{\nu ^{\prime }}%
re^{-\lambda },  \label{three}
\end{equation}
\begin{equation}
kp_t=\frac{e^{-\lambda }}4\left( 2\nu ^{\prime \prime }+\nu ^{\prime 2}+%
\frac{2\nu ^{\prime }}r-\nu ^{\prime }\lambda ^{\prime }-\frac{2\lambda
^{\prime }}r\right) ,  \label{four}
\end{equation}
where $\rho $ is the matter density, $p_r$ is the radial pressure, $p_t$ is
the tangential one, $^{\prime }$ means a radial derivative and 
\begin{equation}
k=\frac{8\pi G}{c^4}.  \label{five}
\end{equation}
Here $G$ is the gravitational constant and $c$ is the speed of light.

The gravitational mass in a sphere of radius $r$ is given by 
\begin{equation}
m=\frac{kc^2}2\int_0^r\rho \left( \omega \right) \omega ^2d\omega .
\label{six}
\end{equation}
Due to $kc^2$, its dimension is length. Then Eq. (2) gives 
\begin{equation}
e^{-\lambda }=1-\frac{2m}r.  \label{seven}
\end{equation}
The compactness of the star $u$ is defined by 
\begin{equation}
u=\frac{2m}r  \label{eight}
\end{equation}
and is dimensionless.

On the other side, the redshift $Z$ depends on $\nu $: 
\begin{equation}
Z\left( r\right) =e^{-\nu /2}-1.  \label{nine}
\end{equation}
The field equations do not contain $\nu $, but its first and second
derivative. One can express $\nu ^{\prime }$ from Eqs. (2,3,7) as 
\begin{equation}
\nu ^{\prime }=\frac{krp_r+2m/r^2}{1-2m/r}.  \label{ten}
\end{equation}
The second derivative $\nu ^{\prime \prime }$ may be excluded by
differentiation of Eq. (3) and combination with the other field equations.
The result is 
\begin{equation}
p_r^{^{\prime }}=-\frac 12\left( \rho c^2+p_r\right) \nu ^{\prime }+\frac{%
2\Delta }r,  \label{eleven}
\end{equation}
where $\Delta =p_t-p_r$ is the anisotropic factor. Combining (10) and (11)
one gets the well-known TOV (Tolman, Oppenheimer, Volkoff) equation \cite
{ten}, \cite{finine} of hydrostatic equilibrium in a relativistic
anisotropic star \cite{five} 
\begin{equation}
p_r^{^{\prime }}=-\left( \rho c^2+p_r\right) \frac{krp_r+2m/r^2}{2\left(
1-2m/r\right) }+\frac{2\left( p_t-p_r\right) }r.  \label{twelve}
\end{equation}
The hydrostatic force on the left $F_h$ is balanced by the gravitational $%
F_g $ and the anisotropic forces $F_a$ on the right. This equation is not
independent from the field equations but is their consequence. It can
replace one of them. This equation is not independent from the field
equations, but is their consequence. It can replace one of them. It is also
equivalent to the Bianchi identities $T_{\nu ;\mu }^\mu =0$, which in the
static spherically symmetric case have only one non-trivial component \cite
{five}, \cite{sefive}, \cite{sesix}, \cite{hufour}. In CGS units $%
G=6.674\times 10^{-8}$ $cm^3/g.s^2$, $c=3\times 10^{10}$ $cm/s$, $%
k=2.071\times 10^{-48}$ $s^2/g.cm$, $kc^2=1.864\times 10^{-27}$ $cm/g$. From
now on we set $G=c=1$, passing to usual relativistic units. Then $k=8\pi $.

As a whole, we have three field equations for five unknown functions: $%
\lambda ,\nu ,\rho ,p_r$ and $p_t$. We can choose freely two of them, but
the model will be physically realistic if a number of regularity, matching
and stability conditions are satisfied too.

\section{Conditions for a physically realistic model}

A comparatively reasonable set of conditions includes

C1. The metric potentials are positive and should be finite and free from
singularities in the star's interior and at the centre should satisfy $%
e^{-\lambda \left( 0\right) }=1$ and $e^{\nu \left( 0\right) }=const$.

C2. Matching conditions. At the surface of the star $r=r_s$ the interior
solution should match continuously to the exterior Schwarzschild solution, 
\begin{equation}
ds^2=\left( 1-\frac{2M}r\right) dt^2-\left( 1-\frac{2M}r\right)
^{-1}dr^2-r^2\left( d\theta ^2+\sin ^2\theta d\varphi ^2\right) .
\label{thirteen}
\end{equation}
This determines the metric at the surface 
\begin{equation}
e^{\nu \left( r_s\right) }=e^{-\lambda \left( r_s\right) }=1-\frac{2M}{r_s}.
\label{fourteen}
\end{equation}
In addition, the radial pressure there vanishes, $p_{rs}=0$. Neither the
energy density nor the tangential pressure are obliged to do so.

C3. The interior redshift $Z$, which, according to Eq. (9), depends only on $%
\nu $ should decrease with the increase of $r$. The surface redshift and
compactness are related, due to Eq. (14): 
\begin{equation}
Z_s=\left( 1-u_s\right) ^{-1/2}-1.  \label{fifteen}
\end{equation}
They should be less than the universal bounds, found when different energy
conditions hold (see C6). In the isotropic case they are $2$ and $8/9$
correspondingly \cite{huseven}. In the anisotropic case, when DEC holds,
they are $5.211$ and $0.974$. When SEC holds, one has the bounds $3.842$ and 
$0.957$ \cite{sixty}. They are greater than those in the isotropic case, but
not arbitrary as asserted in \cite{five}.

C4. The density and the pressures should be non-negative inside the star.
For $\rho $ this coincides with the null energy condition (NEC). At the
centre they should be finite $\rho \left( 0\right) =\rho _0$, $p_r\left(
0\right) =p_{r0}$, $p_t\left( 0\right) =p_{t0}$. Moreover, $p_{r0}=p_{t0}$.

C5. They should reach a maximum at the centre, so $\rho ^{\prime }\left(
0\right) =p_r^{\prime }\left( 0\right) =p_t^{\prime }\left( 0\right) =0$ and
should decrease monotonously outwards, $\rho ^{\prime }\leq 0$, $p_r^{\prime
}\leq 0$, $p_t^{\prime }\leq 0$. The tangential pressure should remain
bigger than the radial one, except at the centre, $p_t\geq p_r$. An
isotropic model is obtained when this inequality turns into an equality,
called the isotropic condition.

C6. Energy conditions. The solution should satisfy the dominant energy
condition (DEC) $\rho \geq p_r$, and $\rho \geq p_t$. When the pressures are
positive, DEC is equivalent to the weak energy condition (WEC). It is
desirable that even the strong energy condition (SEC) $\rho \geq p_r+2p_t$
is satisfied. Obviously, the latter encompasses DEC.

C7. Causality condition. It says that the radial and tangential speeds of
sound should not surpass the speed of light. The speeds of sound are defined
as $v_r^2=dp_r/d\rho $ and $v_t^2=dp_t/d\rho $. Therefore this condition
reads 
\begin{equation}
0<\frac{dp_r}{d\rho }\leq 1,\quad 0<\frac{dp_t}{d\rho }\leq 1.
\label{sixteen}
\end{equation}

C8. The adiabatic index $\Gamma $ as a criterion of stability. This index is
the ratio of two specific heats and should be bigger than $4/3$ for
stability \cite{six}, \cite{hueight}, \cite{hunine}, 
\begin{equation}
\Gamma =\frac{\rho +p_r}{p_r}\frac{dp_r}{d\rho }\geq \frac 43.
\label{seventeen}
\end{equation}

C9. Stability against cracking. Cracking was introduced by Herrera \cite
{huten} as a possibility of breaking of perturbed self-gravitating spheres.
Abreu et al \cite{hueleven} found a simple requirement for avoiding this to
happen, namely the region of stability is 
\begin{equation}
-1\leq \frac{dp_t}{d\rho }-\frac{dp_r}{d\rho }\leq 0.  \label{eighteen}
\end{equation}

C10. The Harrison-Zeldovich-Novikov stability condition \cite{hutwelve}, 
\cite{huthirteen}. It implies that $dM\left( \rho _0\right) /d\rho _0>0$.

\section{General physically realistic solution}

As shown in the introduction, one can choose the two free functions out of
five in a number of ways. However, only the first three conditions in the
previous section are imposed on the metric coefficients. The other concern
the components of the energy-momentum tensor $\rho ,p_r,p_t$. If we choose
two of them we can satisfy many of the above conditions, at least partially,
beforehand and determine the third one through the TOV equation. It does not
contain the metric and replaces Eq. (4) from the original system of Einstein
equations. The TOV equation gives a direct expression for $p_t$%
\begin{equation}
p_t=p_r+\frac 12rp_r^{^{\prime }}+\left( \rho +p_r\right) \frac{\frac
mr+\frac k2r^2p_r}{2\left( 1-\frac{2m}r\right) }.  \label{nineteen}
\end{equation}
The mass $m$ is obtained by integration of the density $\rho $ (see Eq.
(6)), so that the r.h.s. of the above equation involves only $\rho $ and $%
p_r $.

If we try to express $p_r$ from TOV, we get a Riccati equation 
\begin{equation}
R_1p_r^{^{\prime }}=R_2p_r^2+R_3p_r+R_4,  \label{twenty}
\end{equation}
where the coefficients $R_i$ can be easily extracted from Eq. (19). It is
not integrable in the general case.

Eq. (19) is a very complicated integral equation for $\rho $. It simplifies
for $m$ to an Abel equation of the second kind 
\begin{equation}
\left( A_1m+A_2\right) m^{\prime }=A_3m+A_4,  \label{twone}
\end{equation}
where we have used the opposite of Eq. (6), namely 
\begin{equation}
\frac k2\rho =\frac{m^{\prime }}{r^2}.  \label{twtwo}
\end{equation}
The Abel equation does not possess a general solution too.

Thus, the natural step is to choose $\rho $ and $p_r$ as free functions and
express $p_t$ from Eq. (19). Then we find the mass from Eq. (6) and $%
e^\lambda $ from Eq. (7). It replaces the original Eq. (2). The coefficient $%
e^\nu $ is found up to a constant from Eq. (10), which replaces the original
Eq. (3). Hence, instead of the system of Eqs. (2,3,4) we shall use the
system of Eqs. (7,10,19), where $\rho $ and $p_r$ are chosen to satisfy the
parts of conditions C4-C10, referring to them. One can even end with a
single generating function $\rho $ for the solution, by imposing an equation
of state (EOS) on the radial pressure, $p_r=f\left( \rho \right) $.

Conditions C10 concerns only $\rho $, while C8 concerns only $\rho $ and $%
p_r $ and have to be satisfied by properly choosing these two functions.

Another group consists from the conditions upon the metric C1-C3. Eq. (6)
shows that $m\left( 0\right) =\frac mr\left( 0\right) =0$, because $\rho $
is regular and finite at the centre. Eq. (7) states that $e^{\lambda \left(
0\right) }=1$. The mass, being an integral of a positive function, increases
with the radius (which may be seen also from Eq. (22)) and at the surface
Eq. (7) and C2 yield $m\left( r_s\right) =M$, the total mass of the star.
Therefore $m/r$ should increase from $0$ to $M/r_s,$but should not surpass $%
1 $ because a horizon appears then. Thus, $e^\lambda $ is an increasing
function. The r.h.s. of Eq. (10) is positive, hence $\nu ^{\prime }>0$ and $%
e^\nu $ increases too. The arbitrary constant in it allows to arrange the
fulfilment of Eq. (14). Since it shows that $e^{\nu \left( r_s\right) }<1$,
the same is true for $e^{\nu \left( 0\right) }$ and the constant in C1 is
smaller than unity. Finally, $\nu ^{\prime }>0$ means that $Z$ decreases
outwards (see Eq. (9)). The surface redshift and the compactness will
satisfy the bounds in \cite{sixty}, as long as the DEC or the SEC holds.
Thus, conditions C1-C3 are satisfied from general considerations and the
pre-arranged requirement $\left( m/r\right) ^{\prime }>0$.

The third group of requirements consists of the parts of conditions C4-C10,
which refer to $p_t$. One of them follows immediately from Eq. (19). At $r=0$
we have $p_{t0}=p_{r0}$ because $p_r^{^{\prime }}\left( 0\right) =0,\rho
\left( 0\right) $ and $p_r\left( 0\right) $ are finite, $\frac mr\left(
0\right) =0$ as shown above. Thus, $p_{t0}$ is also finite and the solution
satisfies C4 completely. There remain parts of C5, C6, C7 and C9 to be
satisfied by $p_t$.

The usual method used in the literature is to take two of the five essential
characteristics of the model and choose simple expressions for them as
polynomials or rational functions. Sometimes trigonometric and even special
functions are used. They are supplied with enough parameters to try to
satisfy the realistic conditions, starting with C1 and ending with C10.
Already at this stage some of the expressions are so complicated that the
authors pass to graphical description proofs, giving figure after figure.
Even with one free parameter the graphics become 3D. This method is rather
close to numerical simulations and not to an analytical study. Is it
possible to reduce the number of graphic proofs? Do any relations exist
between the numerous conditions, concerning just a few basic characteristics
of the model, so that some of the conditions follow from the others in the
general case? These questions will be dealt with in the next section.

\section{Relations between the different conditions}

Let us take the anti-cracking condition C9 and write it as 
\begin{equation}
-1+\frac{dp_r}{d\rho }\leq \frac{dp_t}{d\rho }\leq \frac{dp_r}{d\rho }.
\label{twthree}
\end{equation}
Since the radial speed of sound from the causality condition is arranged to
lie in the interval $(0,1]$, the l.h.s. is not positive. Combining
inequalities (16) and (23) we get 
\begin{equation}
0\leq \frac{dp_t}{d\rho }\leq \frac{dp_r}{d\rho }.  \label{twfour}
\end{equation}
This can be written as 
\begin{equation}
0\leq \frac{p_t^{^{\prime }}}{\rho ^{\prime }}\leq \frac{p_r^{\prime }}{\rho
^{\prime }}.  \label{twfive}
\end{equation}
Multiplying by $\rho ^{\prime }$, which should be negative, results in the
pair of inequalities 
\begin{equation}
0\geq p_t^{\prime }\geq p_r^{\prime }.  \label{twsix}
\end{equation}
Starting with Eq. (26) we may go backwards to Eq. (24). It can be replaced
by the second part of Eq. (16), which gives a weaker pair of inequalities.
Eq. (24) may be replaced also by the weaker pair given by Eq. (23), which is
nothing but Eq. (18). Thus we have proved that Eq. (26) is equivalent to C7
and C9 and if it holds, they hold too.

We take Eq. (26) as the basic one that we have to satisfy. It shows that $%
p_t $ and $p_r$ decrease monotonously (for $p_r$ this is arranged
beforehand) and that $p_t^{\prime }\left( 0\right) =0$ as long as $%
p_r^{\prime }\left( 0\right) =0$. The latter equality is also pre-arranged.
All these enter C5, together with the pre-arranged conditions for $\rho $.
C5 will be satisfied completely when we prove that $p_t\geq p_r$. Eq. (26)
also shows that $\Delta $ increases with $r$.

In order to do this proof we first study the behaviour of inequalities under
differentiation and integration. If $g\left( r\right) \geq 0$, $g^{\prime
}\left( r\right) \geq 0$ does not necessarily follow, because the function $%
g\left( r\right) $ may oscillate up and down, having regions with negative
derivative, but still remaining positive all the time. However, if $g\left(
r\right) \geq 0$, the definite integral of it $\int_{r_1}^{r_2}g\left(
r\right) dr\geq 0$ too, provided that $r_2>r_1$, because this is the area of
the surface under the graphic of $g\left( r\right) $. The analogous
conclusion also holds for a negative $g\left( r\right) $. There are two
special radial points in a star model: $r=0$ and $r=r_s$. So we shall take
integrals between $0$ and $r$, or between $r$ and $r_s$. These lead to
corollaries of the initial inequality and not to equivalence. The easiest
function to integrate is a derivative, $g=h^{\prime }$. This is one of the
reasons to start from the end of the list of conditions, where derivatives
prevail the inequalities, and work backwards, contrary to what is usually
done.

Thus, taking the integral $\int_0^r$ of Eq. (26) we obtain 
\begin{equation}
0\geq p_t-p_{t0}\geq p_r-p_{r0}.  \label{twseven}
\end{equation}
We have shown earlier that TOV leads to $p_{t0}=p_{r0}$, hence, Eq. (27)
becomes $p_{r0}\geq $ $p_t\geq p_r$. In this way C5 holds in total, provided
that Eq. (26) is true.

It remains to satisfy C6, that is, the two energy conditions. Let's deal
first with DEC. The first part of C7 may be written as 
\begin{equation}
\left( p_r-\rho \right) ^{\prime }\geq 0.  \label{tweight}
\end{equation}
Integrating this inequality from $r$ to $r_s$ and taking into account that $%
p_{rs}=0$, we find 
\begin{equation}
\rho -p_r\geq \rho _s\geq 0.  \label{twnine}
\end{equation}
Thus, DEC follows for $p_r$ from the causality condition. In fact, we can
pre-arrange both of them. Doing the same procedure for $p_t$ yields a
similar result 
\begin{equation}
\rho -p_t\geq \rho _s-p_{ts}  \label{thirty}
\end{equation}
Hence, if DEC holds at the surface, it also holds in the interior. Now, at
the surface the expression for $p_t$ supplied by Eq. (19) becomes 
\begin{equation}
p_{ts}=\frac 12r_sp_r^{^{\prime }}\left( r_s\right) +\frac{\rho _s\frac
M{r_s}}{2\left( 1-\frac{2M}{r_s}\right) }.  \label{thone}
\end{equation}
Then $\rho _s-p_{ts}\geq 0$ turns into 
\begin{equation}
\frac{2-\frac{5M}{r_s}}{2\left( 1-\frac{2M}{rs}\right) }\rho _s\geq \frac
12r_sp_r^{^{\prime }}\left( r_s\right) .  \label{thtwo}
\end{equation}
The r.h.s. is negative while the denominator in the l.h.s. is positive.
Consequently, a sufficient condition for DEC will be a positive numerator, 
\begin{equation}
u_s=\frac{2M}{r_s}\leq 0,8.  \label{ththree}
\end{equation}
It is a rather realistic condition.

Let us discuss next SEC. A sufficient condition, which depends only on $\rho 
$ and $p_r$ and therefore can be pre-arranged, reads 
\begin{equation}
2p_{r0}+p_r\leq \rho .  \label{thfour}
\end{equation}
Then SEC follows from the chain of inequalities 
\begin{equation}
2p_t+p_r\leq 2p_{t0}+p_r=2p_{r0}+p_r\leq \rho ,  \label{thfive}
\end{equation}
which are due to the fact that $p_t$ decreases and is equal to $p_r$ at the
origin. Eq. (34) is also a rather mild restriction.

Summarizing, the pair of inequalities (26) are the only conditions that $p_t$
must satisfy. Before discussing them we present a relation about C8. A
sufficient condition for it to take place is a lower limit on the radial
speed of sound 
\begin{equation}
\frac{dp_r}{d\rho }\geq \frac 13,  \label{thsix}
\end{equation}
which is simpler. Indeed, when SEC (C6) holds and we apply C5 we get $\rho
\geq 3p_r$. Put in C8 it gives the above inequality.

Now let us take Eq. (27) and place in it the expression for $p_t$ from Eq.
(19). After some manipulations we get 
\begin{equation}
-rp_r^{^{\prime }}\leq 2F\leq -rp_r^{^{\prime }}+2\left( p_{r0}-p_r\right)
,\quad 2F=r^2\left( \rho +p_r\right) \frac{\frac m{r^3}+\frac k2p_r}{1-\frac{%
2m}r}.  \label{thseven}
\end{equation}
Applying the L'Hospital rule to Eq. (6) yields 
\begin{equation}
\frac mr\left( 0\right) =\frac m{r^2}\left( 0\right) =0,\quad \frac
m{r^3}\left( 0\right) =\frac k6\rho _0.  \label{theight}
\end{equation}
Therefore, $F$ behaves as $r^2$ when $r\rightarrow 0$. The left and the
right bounds should behave the same way, otherwise one of the inequalities
will break near the origin. This means that $p_r\sim p_{r0}-c_1r^2$, $%
p_r^{\prime }\sim -2c_1r$ there. Hence if we choose $p_r=p_{r0}-c_1r^n$ as a
simple expression for the radial pressure, only $n=2$ may fulfil Eq. (27).
If $p_r$ satisfies an EOS, we have $p_r^{\prime }=f\left( \rho \right) _\rho
\rho ^{\prime }\sim r$ and $\rho \sim \rho _0-c_1r^2$. If $\rho =\rho
_0-ar^n $ \cite{fisix}, only $n=2$ has a chance. Here $c_1,a,n$ are some
positive constants.

\section{Solutions with linear EOS and simple energy density}

Let us choose $\rho $ as a simple binomial and $p_r$ satisfying a linear EOS
with a bag constant 
\begin{equation}
\rho =\rho _0-ar^2,  \label{thnine}
\end{equation}
\begin{equation}
p_r=\beta \left( \rho -\rho _s\right) .  \label{forty}
\end{equation}
This kind of density appears already in the Tolman VII isotropic solution 
\cite{ten}. Anisotropic models with such density include \cite{fofive}, \cite
{fisix}, \cite{siseven}, \cite{sieight}, \cite{sinine}, \cite{seventy}, \cite
{eitwo}, \cite{nifive}, \cite{ninine}, \cite{hundred}, \cite{hutwo}, \cite
{hufive}, \cite{husix}. Models with such LEOS are discussed in \cite{sifive}%
, \cite{nithree}, \cite{nifour}, \cite{nifive}, \cite{nisix}, \cite{niseven}%
, \cite{nieight}, \cite{ninine}. Obviously $p_{r0}=\beta \left( \rho _0-\rho
_s\right) >0,$ $p_{rs}=0$. Thus C4 holds for them. The $\rho $ and $p_r$
have a maximum at the centre and decrease outwards, satisfying their part of
C5. We also introduce the notation 
\begin{equation}
x=r^2,\quad y=\frac x{x_s},\quad \alpha =1-\frac{\rho _s}{\rho _0},\quad
b=k\rho _0x_s.  \label{foone}
\end{equation}

Now $dp_r/d\rho =\beta $, so the first causality condition in C7 holds for $%
0<\beta \leq 1$. Then DEC holds too. The sufficient condition for SEC (34)
becomes 
\begin{equation}
\beta \left( 2\rho _0-3\rho _s\right) \leq \left( 1-\beta \right) \rho .
\label{fotwo}
\end{equation}
This inequality is true, if it is true for $\rho =\rho _s$, which yields 
\begin{equation}
\alpha \leq \frac 1{1+2\beta }.  \label{fothree}
\end{equation}

Next we can obtain a more refined estimate for the adiabatic index $\Gamma $
and C8. Inequality (17) leads to 
\begin{equation}
\left( 3\beta -1\right) \rho \geq \left( 3\beta -4\right) \rho _s.
\label{fofour}
\end{equation}
It is true when $\beta \geq 1/3$ since the r.h.s. is always negative,
because $\beta \leq 1$, while the l.h.s. is non-negative. Hence, C8 holds,
in accord with the sufficient condition Eq. (36). When $\beta <1/3$ we get 
\begin{equation}
\rho <\left( 1+\frac 3{1-3\beta }\right) \rho _s,  \label{fofive}
\end{equation}
which holds, provided it holds for $\rho _0$. This leads to $\alpha
<3/\left( 4-3\beta \right) $. Then C8 is also fulfilled.

Eq. (39) yields for $a$%
\begin{equation}
a=\frac{\rho _0-\rho _s}{x_s}=\frac{\alpha \rho _0}{x_s}.  \label{fosix}
\end{equation}
Then $\rho ,p_r$ may be written as 
\begin{equation}
\rho =\rho _0\left( 1-\alpha y\right) ,\quad p_r=\alpha \beta \rho _0\left(
1-y\right) .  \label{foseven}
\end{equation}
The mass $m$ is given by Eq. (6) 
\begin{equation}
m=\frac{k\rho _0}{30}r^3\left( 5-3\alpha y\right) ,\quad m_s=M=\frac{k\rho _0%
}{30}r_s^3\left( 5-3\alpha \right) ,\quad u_s=\frac b{15}\left( 5-3\alpha
\right) .  \label{foeight}
\end{equation}
As a function of $\rho _0$, the total mass $M$ and its derivative become 
\begin{equation}
M\left( \rho _0\right) =\frac{kr_s^3}{30}\left( 2\rho _0+3\rho _s\right)
,\quad \frac{dM\left( \rho _0\right) }{d\rho _0}=\frac{kr_s^3}{15}>0.
\label{fonine}
\end{equation}
Thus C10 is satisfied and these specific density and radial pressure satisfy
all realistic conditions.

Let us turn to the metric functions next. Putting Eq. (48) into Eq. (7)
provides an expression for $\lambda $%
\begin{equation}
e^{-\lambda }=1-\frac b{15}y\left( 5-3\alpha y\right)  \label{fifty}
\end{equation}
This is a polynomial of second degree in $x$ \cite{nieight}. We have pointed
out that $m/r$ should increase with $r$ from $0$ to $M/r_s$. Eq. (48) gives 
\begin{equation}
\left( \frac mr\right) ^{\prime }=\frac{k\rho _0r}{15}\left( 5-6\alpha
y\right) .  \label{fione}
\end{equation}
This will be positive when $\rho _0<6\rho _s$, equivalent to $\alpha <5/6$.

The function $e^\nu $ is found by integrating Eq. (10) which is a rational
function in $y$. The result is given by Eq. 3.4 from \cite{nieight}. These
functions satisfy the conditions on the metric C1-C3, as was shown before.

\section{Tangential pressure and the main pair of inequalities}

Eq. (19) may be rewritten as 
\begin{equation}
p_t=\left( xp_r\right) _x+F,  \label{fitwo}
\end{equation}
\begin{equation}
F=\frac{x\left( \rho +p_r\right) \left( \frac m{r^3}+\frac k2p_r\right) }{%
2\left( 1-\frac{2m}r\right) }.  \label{fithree}
\end{equation}
The inequalities (26) become 
\begin{equation}
p_{rx}-\left( xp_r\right) _{xx}\leq F_x\leq -\left( xp_r\right) _{xx}.
\label{fifour}
\end{equation}
Using the equations in the previous section we get 
\begin{equation}
2\alpha \beta \rho _0\leq 2F_y\leq 4\alpha \beta \rho _0.  \label{fifive}
\end{equation}
The function in the middle should lay between two simple constant bounds. It
is given by 
\begin{equation}
2F=\frac{b\rho _0}{30}\frac{yF_1F_2}{F_3},  \label{fisix}
\end{equation}
where 
\begin{equation}
F_1=1+\alpha \beta -\alpha \left( 1+\beta \right) y,  \label{fiseven}
\end{equation}
\begin{equation}
F_2=5+15\alpha \beta -3\alpha \left( 1+5\beta \right) y,  \label{fieight}
\end{equation}
\begin{equation}
F_3=1-\frac b{15}y\left( 5-3\alpha y\right) ,  \label{finine}
\end{equation}
all of them being positive. Next we find 
\begin{equation}
2F_y=\frac{b\rho _0}{30F_3^2}\left\{ \left[ F_1F_2-\alpha \left( 1+\beta
\right) yF_2-3\alpha \left( 1+5\beta \right) yF_1\right] F_3+\frac
b{15}F_1F_2y\left( 5-6\alpha y\right) \right\} .  \label{sixty}
\end{equation}
Eq. (12) shows that $F$ is closely connected to the gravitational force in
the TOV equation, $F_g=-2F/r$. We expect that $F_y$ decreases. Then it
satisfies Eq. (55) when it does so for $r=0$ and $r=r_s$.

At $r=0,$ $y=0$ and we have 
\begin{equation}
\frac{12\gamma }{\left( 1+\gamma \right) \left( 1+3\gamma \right) }\leq
b\leq \frac{24\gamma }{\left( 1+\gamma \right) \left( 1+3\gamma \right) },
\label{sione}
\end{equation}
where $\gamma =\alpha \beta $.

At the surface $r_s,$ $y=1$ and 
\begin{equation}
4\gamma l^2\leq \left[ 9\alpha ^2-16\alpha +5-2\gamma \left( 10-9\alpha
\right) \right] l+\left( 1-\alpha \right) \left( 5-3\alpha \right) \left(
5-6\alpha \right) \leq 8\gamma l^2,  \label{sitwo}
\end{equation}
where we have replaced $b\,$ with $l$ for simplicity, 
\begin{equation}
b=\frac{15}{l+5-3\alpha }.  \label{sithree}
\end{equation}

The final result is four intricate inequalities for the three free constants 
$\alpha ,\beta ,b$. Eq. (41) shows that they hide inside the physical
constants $\rho _0,\rho _s$ and $r_s$, from which through Eq. (48) one can
find the other main star characteristics $M,u_s,Z_s$. In addition, $\alpha $
should satisfy Eqs. (43,51), while the surface compactness - Eq.(33).

\begin{figure}
   \centering
      \includegraphics[scale=.8]{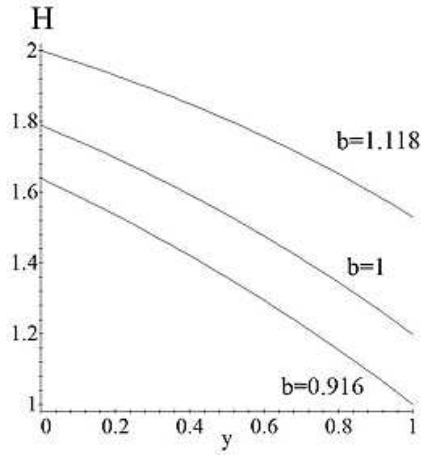}
   \caption{The decrease of $H$ for the region of Eq.(66)}
\end{figure}
A solution with the same functions $\rho $ and $p_r$ was given in \cite
{nieight}. It has $\rho _0=3.98\times 10^{15}$ $g/cm^3$, $\rho _s=3.29\times
10^{15}$ $g/cm^3$ and $\beta =1/3$. Then we have $\alpha =0.1734$, $\gamma
=0.0578$. Obviously $\alpha $ satisfies Eqs. (43,51), while the radial speed
of sound satisfies the sufficient condition Eq. (36). Eq. (61) becomes 
\begin{equation}
0.559\leq b\leq 1.118.  \label{sifour}
\end{equation}
Eq. (62) yields a region which intersects with it, namely 
\begin{equation}
0.916\leq b\leq 1.251,  \label{sifive}
\end{equation}
so that, finally 
\begin{equation}
0.916\leq b\leq 1.118  \label{sisix}
\end{equation}
is the interval for a realistic $b$. We can also show graphically that $%
H\equiv F_y/\gamma \rho _0$ decreases for the interval (66), as required,
and stays between $1$ and $2$, because of Eq.55, see Fig.1. The middle line
is for $b=1$, while the lower and the upper ones correspond to the limits in
Eq.(66). This is the only time we use a graphic proof.

In \cite{nieight} $b=1.074$, which is in the above region. One obtains also $%
u_s=0.299b=0.321$, a value well below $0.8$ and the bounds in \cite{sixty}.
Therefore, such a solution satisfies C1-C10 and is physically realistic. The
star is chosen very compact, with radius $3.8$ km. The solar mass is $%
M_{sol}=1.988\times 10^{33}$ g, so that $GM_{sol}/c^2=1.474$ km. Then the
mass is $M=0.411M_{sol}$ and $Z_s=0.2135$. We have found instead a range of
solutions, with $b$ satisfying Eq. (66), which include the solution of \cite
{nieight}. The main parameters also span ranges of values. Thus $u_s\in
\left[ 0.274,0.334\right] $, $M/M_{sol}\in \left[ 0.353,0.431\right] $, $%
Z_s\in \left[ 0.174,0.225\right] $.

Astronomers have observed even more compact, presumably quark stars, such as
PSR B0943+10 with radius $2.6$ km, described also by an analytical model 
\cite{huone}, Table 1 and \cite{hutwo}, Table 1. Another observed star
mentioned in \cite{hutwo}, named RX J1856.5-3754, has radius of $3.5$ km, so
the above value of $3.8$ km is realistic.

Next, let us find another cluster of solutions by choosing $\rho _s/\rho
_0=0.82$, ($\alpha =0.18$), $\beta =1/3$. Then $\gamma =0.06.$ Eqs. (61,62)
this time lead to 
\begin{equation}
0.958\leq b\leq 1.152,  \label{siseven}
\end{equation}
which is a slightly displaced range. Once again $H$ decreases in it, staying
in the interval $[1,2]$, see Fig.2. We have $u_s\in \left[
0.285,0.342\right] $, $M/M_{sol}\in \left[ 0.367,0.441\right] $, $Z_s\in
\left[ 0.183,0.233\right] $.

\begin{figure}
   \centering
      \includegraphics[scale=.8]{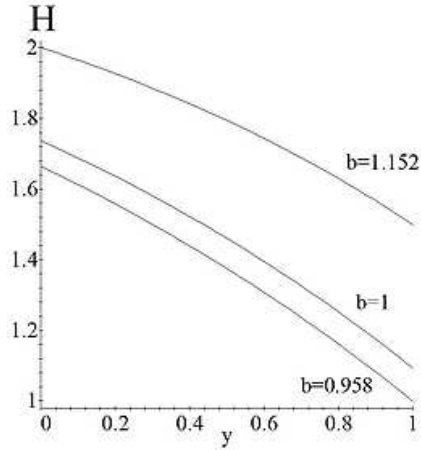}
   \caption{The decrease of $H$ for the region of Eq.(67)}
\end{figure} 
If we take a constant density solution, $\rho =\rho _0=\rho _s$, we'll have $%
\alpha =0$ and $u_s=0.333b$, which is close to the values, discussed above.
This is not surprising, since the density of the crust is taken high, more
than 80\% of the central density. However, this solution is unphysical,
since the speed of sound in both directions is infinite. The solutions above
satisfy C1-C10 and are physically realistic. There is an interesting
relation, following from that. Eq. (41), written in kind of CGS units,
yields 
\begin{equation}
b=kc^2\rho _0r_s^2=1,864.10^{-17}\rho _0r_s^2,  \label{sieight}
\end{equation}
when $\rho _0$ is measured in $g/cm^3$, while $r_s$ is measured in km. It
shows that there are models with different radius, but the same $b$. Let us
choose $\rho _0=\rho _1\times 10^{15}$ $g/cm^3$ where $\rho _1$ is a number
between $1$ and $10$. The previous equation becomes 
\begin{equation}
\rho _1r_s^2=53.648b.  \label{sinine}
\end{equation}
Let us take the highest $b$ found in the two examples, $b=1.152$. Then the
above equality holds for a very compact quark star if $r_s=4$ km and $\rho
_1=3.863$. However, it also holds for a neutron star with $r_s=7.861$ km and 
$\rho _1=1$. In both cases the $\rho _0$ is of order $10^{15}$ $g/cm^3$. Eq.
(48) shows that $u_s$ stays constant with $b$, hence, the total mass $M$
should increase when $r_s$ increases. One can call such solutions $b$%
-isotopic.

\section{Some other EOS}

We have already pointed out that models with constant $\rho $ are not
physical, because the speed of sound becomes infinite. Conditions C7-C9
become singular and without sense. The first interior solution was such a
model \cite{two} and others followed \cite{three}, \cite{five} \cite{fiseven}%
, \cite{fieight}, \cite{sione}, \cite{sitwo}, \cite{sithree}, \cite{sesix}.
They can serve as an approximation to some features of the stars.

The same may be said for models with $p_r=0$ and any $\rho $ \cite{three}, 
\cite{fiseven}, \cite{fieight}, \cite{sifour}. Because of C4 we should have $%
p_{t0}=0$, and because of C5 $p_t^{\prime }\leq 0$. Since, according to C4,
the pressures are non-negative, $p_t=0$ throughout the bulk too. What
remains is a dust model with only $\rho \neq 0$. Such models are unstable
and collapse.

Let us discuss models with $\rho $ given by Eq. (39) and linear EOS, but
without the MIT bag constant,\cite{eitwo}, \cite{eithree}, \cite{eifour},
i.e. 
\begin{equation}
p_r=\beta \rho .  \label{seventy}
\end{equation}
They follow from Eq. (39) when $\rho _s=0$ ($\alpha =1$). The energy density
vanishes at the surface, so such stars are gaseous and have no crust. The
tangential pressure is given again by Eq. (52). Now 
\begin{equation}
\rho +p_r=\left( 1+\beta \right) \rho _0\left( 1-y\right) .  \label{seone}
\end{equation}
Then $F\sim y\left( 1-y\right) $ and vanishes at $r=0$ ($y=0$) and $r=r_s$ ($%
y=1$). Further, 
\begin{equation}
\left( xp_r\right) _x=\beta \rho _0\left( 1-2y\right) .  \label{setwo}
\end{equation}
Thus 
\begin{equation}
p_{t0}=\beta \rho _0,\quad p_{ts}=-\beta \rho _0,  \label{sethree}
\end{equation}
that is, the tangential pressure changes sign and becomes negative. This is
not allowed by C4 and the model is unphysical with such energy density.

Our final example are models with the same $\rho $, but with polytropic EOS 
\cite{nifive} 
\begin{equation}
p_r=K\rho ^{1+\frac 1n},  \label{sefour}
\end{equation}
where $K$ is the polytropic constant and $n$ is the polytropic index. Once
again $\rho _s=0$ and $\alpha =1$. Then Eq. (39) leads to 
\begin{equation}
\rho =\rho _0\left( 1-y\right) ,\quad p_r=p_{r0}\left( 1-y\right) ^{1+\frac
1n},  \label{sefive}
\end{equation}
\begin{equation}
\rho +p_r=\left( 1-y\right) \left[ \rho _0+p_{r0}\left( 1-y\right) ^{\frac
1n}\right]  \label{sesix}
\end{equation}
and again $F\sim y\left( 1-y\right) $. On the other side 
\begin{equation}
xp_{rx}=-\left( 1+\frac 1n\right) p_{r0}y\left( 1-y\right) ^{\frac 1n}.
\label{seseven}
\end{equation}
Then 
\begin{equation}
\Delta =-\left( 1+\frac 1n\right) p_{r0}y\left( 1-y\right) ^{\frac 1n}+F
\label{seeight}
\end{equation}
and vanishes at the centre and at the surface of the star. Thus $\Delta _x$
changes sign somewhere. However, C5 obliges $\Delta _x$ to be positive so
that $\Delta $ increases. Hence, this model is unphysical, too. One should
change the simple expression for $\rho $. These examples show that it is not
so easy to find physically realistic star models.

\section{Discussion}

With five unknown functions and three equations, one has 10 combinations of
two free functions, claiming to give the general solution for anisotropic
static stars. The panorama given in the Introduction shows that most of them
have been applied to find at least one concrete solution. Instead of $p_t$
the anisotropic factor $\Delta $ has been used. Sometimes $m$ is taken
instead of $\rho $ or $p_r$. The general solution was seriously discussed in 
\cite{theight}. The main shortcoming is that the conditions for a realistic
model are checked after the choice is made. The expressions for the
different characteristics become very involved even for polynomial seeding
functions and one has to turn to graphic descriptions, which serve as
proofs. Solutions are usually supplied with lots of constants, to negotiate
the set C1-C10. Just one constant is enough to turn a 2-dimensional plot
into a 3-dimensional one, whose projection as a 2-dimensional flow of lines
is often used. With two and more constants only partial plots are possible.

We have argued that the combination of free functions $\rho $, $p_r$ seems
to be the right choice to reduce the number of graphic proofs. There is a
simple chain of relations between $\rho $, $m$, and $\lambda $, due to Eqs.
(6,7), so that one may start with some of the last two and $p_r$. The
advantage of this solution generation method is that we can arrange
beforehand that the part of the condition set, referring to $\rho $, and $%
p_r $, is satisfied. This concerns the majority of conditions, namely
C4-C10. Besides, C8 and C10 contain only $\rho $ and $p_r$.

The set of conditions C1-C10 was formed during the years. In older papers
just the new solution is given and sometimes its matching to the exterior.
Later the conditions for monotonous decrease and the energy conditions were
added. Then followed the causality and stability conditions, the last being
C10, which, however dates back to 1965. These conditions come from
phenomenology, where hydrostatics, nuclear theory and thermodynamics meet
with general relativity. In theoretical astrophysics they stand as axioms
that must be satisfied. One is tempted to ask whether all these conditions
are independent. We have shown that this is not the case and reduced them to
the couple of inequalities in Eq. (26). Only Eq. (26) needs a graphic proof
in the concrete examples. To emphasize this point we haven't given any other
graphics, even for illustration purposes. This reduction is possible due to
the following factors.

The characteristics of static models, unlike the dynamical ones, depend on
one variable, the radius. This is true even for the stability criteria,
based on time-dependent perturbations. Thus, there are only ordinary
derivatives and all of them can be replaced by $r$-derivatives, e.g. $%
dp_r/d\rho =p_r^{\prime }/\rho ^{\prime }$.

Inequalities may be integrated, obtaining other true inequalities, which are
corollaries, not equivalent to their parents.

Most of the features of the model have simple monotonous behaviour as
functions of the radius. Thus, $\rho $, $p_r$, $p_t$, $Z$ decrease, while $%
\lambda $, $\nu $, $m$, $\Delta $, $u$ increase with $r$. Therefore, one can
multiply the inequalities with their derivatives, changing the sign when
necessary, without bothering about the region of application - it remains
the same. When there is a bump (mainly in $p_t$), this signals that
something is wrong with the causality and/or the anti-cracking conditions.

Thus, many of the conditions follow from the others for any solution. When
an EOS is specified, some of the conditions are fulfilled, provided only $%
\rho _s$ and $\rho _0$ are known and not the whole graphic of $\rho $.

To illustrate this general formalism we have given a solution with simple
energy density and simple EOS, namely linear EOS with bag constant. We have
found that the general solution depends on three free constants. They hide
inside themselves the physical constants $\rho _0$, $\rho _s$, $p_{r0}$ and $%
r_s$, which lead in turn to other main characteristics like $M$, $u_s$, and $%
Z_s$, some of which are measured in astronomy. We have found regions where
the solution is realistic and not just a point in constants' space. In the
previous section we have shown why some popular in the past solutions break
certain realistic conditions. This is true even for the $\gamma $-law and
the polytropic EOS with the same simple density. Hence, other expressions
for $\rho $ should be studied. We hope that the application of the method,
described in this paper, will lead to many other solutions in the future,
obtained in an easier study of genuine analytic nature.

\end{document}